# Nature of Magnetoelectric coupling in corundum antiferromagnet $Co_4Ta_2O_9$

S. Chaudhary, P. Srivastava and S. Patnaik

*School of Physical Sciences, Jawaharlal Nehru University, New Delhi-110067, India*

**Abstract**

We study the magnetocapacitance (MC) effect and magnetoelectric (ME) coupling in spin-flop driven antiferromagnet $Co_4Ta_2O_9$. The magnetocapacitance data at high magnetic fields are analyzed by phenomenological Ginzburg-landau theory of ferroelectromagnets and it is found that change in dielectric constant is proportional to the square of magnetization. The saturation polarization and magnetoelectric coupling are estimated to be $52 \mu C/m^2$ and $\gamma = 1.4 \times 10^{-3}$ $(emu/g)^{-2}$ respectively at 6 Tesla. Electric polarization is achieved below Neel temperature only when the sample is cooled in the presence of magnetic field and it is established that the ground state is non-ferroelectric implying that magnetic lattice does not lead to spontaneous symmetry breaking in $Co_4Ta_2O_9$.

## I. Introduction

The magnetic control of an electrically ordered state or electric control of a magnetically ordered state promises to usher in a plethora of advanced technologies [1-2] along with the possibility of finding new functionalities hitherto unknown. At the center of such juxtaposed phenomena is the magneto-electric (ME) effect that was discovered over a century ago by Curie.[3] However, the strength of ME coupling had remained extremely weak until its prediction in antiferromagnetic $Cr_2O_3$ by Dzyaloshinskii.[4] Over the years, several insulating magnetic oxides have been identified with strong ME coupling and a system of recent interest includes corundum of the general formula $Z_4A_2O_9$ (with Z= Co, Fe, Mn, and A = Nb, Ta).[5] In particular, two members of this family, $Co_4Nb_2O_9$ and $Co_4Ta_2O_9$ have attracted sustained scientific scrutiny. The identification of microscopic origin of such codependence that relates to spatial inversion and time reversal symmetry breaking is of current interest.

Experimentally, the ME coupling is estimated from the magnetic field dependence of the dielectric constant (so called Magnetocapacitance) and based on this, mutiferroic compounds are grouped into two sets. One set belongs to the case where the magnetic structure drives the onset of ferroelectric phase (e.g. $TbMnO_3$, $GdFeO_3$ etc) that does not require poling in the presence of external field.[6-8] The second group shows linear ME effect (e.g. $Cr_2O_3$, $MnTiO_3$, $NdCrTiO_5$, $Co_3O_4$ etc) where presumably the spontaneous electric polarization is absent at the ground state.[5,6,9] Only when an external magnetic field is applied during cooling, the electric polarization is developed, the magnitude of which increases linearly with field.[10,11,12] Further, such co-dependence are widely studied under the general framework of Ginzburg–Landau theory for second-order phase transition of ferroelectromagnets.[13-17]

Recently, Fang et al. have reported that electric polarization can be induced in the $Co_4Ta_2O_9$ only under the condition that the sample is poled in the presence of magnetic field.[18] But the cause of this ME effect in $Co_4Ta_2O_9$ and its ground state property has remained unknown. Moreover, magnetic spin-flop transition in $Co_4Ta_2O_9$ has been indicated but the magnitude of ME effect is not estimated. In this paper, we report a detailed investigation of magnetoelectric effect in polycrystalline $Co_4Ta_2O_9$ that includes (i) determination of Magneto-electric (ME) coefficient, (ii) correlation between magnetism and magneto-capacitance in $Co_4Ta_2O_9$ (iii) the effect of spin flop transition on large ME coupling and (iv) establishment of non-ferroelectric ground state in $Co_4Ta_2O_9$.

## II. Experiments

The polycrystalline $Co_4Ta_2O_9$ samples were prepared by solid state reaction method. Stoichiometric amount of $Co_3O_4$ and $Ta_2O_5$ were used as starting material. The mixture was ground together and calcined at 1000°C for 10 hrs in air. The powder was then reground and pressed into pellets of diameter 10mm and 1mm thickness and sintered at 1100°C for another 10 hrs. Both heating and cooling rates were kept at 5 K/min. Phase purity was confirmed at room temperature by powder X-ray diffraction (Rigaku Miniflex X-ray diffractometer with Cu-$K_\alpha$ radiation). Lattice parameters were obtained from Reitveld refinement of XRD data using FullProf software. The DC magnetization measurements were done in vibrating sample magnetometer (VSM) mode of a *Cryogenic* 14 Tesla Physical Property Measurement System (PPMS). The dielectric measurements were done using an Agilent E4980A LCR meter. For pyroelectric and dielectric measurements, electrodes were prepared on the sample by painting silver paste on the planar surfaces. The pyroelectric measurement was performed using a Kiethley 6514 electrometer and polarization was obtained by integrating pyroelectric current over time. The sample was poled each time from well above ordering temperature under various combinations of electric and magnetic field. The poling field was removed at lowest temperature and terminals were sorted for 15 minutes to avoid any role of electrostatic stray charges. The warming ramp rate was fixed at 5 K/min.

## III. Results and discussion

Room temperature powder X-ray diffraction (XRD) pattern of polycrystalline $Co_4Ta_2O_9$ sample in the range of $2\theta = 20°$ to $80°$ is shown in Fig.1(a). The refinement done by Fullprof technique has been used for phase detection of synthesized sample. It is confirmed that all the spectra fit the standard database for the $\alpha$-$Al_2O_3$-type trigonal type structure with space group $P\bar{3}c1$. From the XRD data, the lattice parameters are estimated to be a = 5.1714 Å, c = 14.1451 Å with $\alpha$=90.00° and, $\gamma$ =120.00°. X-ray diffraction data confirm almost phase pure synthesis of polycrystalline $Co_4Ta_2O_9$. The atomic coordinates are also estimated from the room temperature powder XRD and are summarized in Table 1. The $\alpha$-$Al_2O_3$-type structure for $Co_4Ta_2O_9$ is schematically illustrated in Fig. 1(b). It is seen that two crystallographic sites for the Co ions are non-equivalent. The unit cell consists of two formula units of $Co_4Ta_2O_9$. Such corundum type of structure of general formula $Z_4A_2O_9$, were first reported by Bertaut et al..[5]

Scanning electron microscopic (SEM) photographs of $Co_4Ta_2O_9$ are shown in Fig. 2(a). It is evident that the compound is dense and average grain size is large. In order to confirm the phase purity of as grown sample, energy dispersive X-ray analysis (EDX) was carried out for synthesized compound (Fig 2(b)). No impurity element was observed. The temperature dependence of magnetization for $Co_4Ta_2O_9$ at various external magnetic fields (0.01, 0.1, 1, 1.5, 2 T) is shown in Fig. 3(a). Under a magnetic field of 0.01 T and zero field cooled warming protocol, it is observed that the magnetization first increases linearly with increasing temperature. Above $T_N$ = 20 K, the magnetization starts to drop upon warming further, showing a typical antiferromagnetic (AFM) nature. The temperature corresponding to peak in magnetization is identified as the Ne´el temperature ($T_N$). This is determined more accurately from maximum in the dM/dT curve (inset of Fig. 3 (a)). Similar trend is also observed as the external field is increased to 0.1 T. However, when the external magnetic field exceeds 0.9 T, the magnetization curves show a distinctly different behavior; the magnetization decreases with increasing temperature from the lowest measured temperature with a slight deviation around $T_N$. This is a clear signature of spin-flop transition for antiferromagnetic $Co_4Ta_2O_9$.[19,20] Drawing analogy with $Co_4Nb_2O_9$ [20] we can conclude that a large magnetic field applied along the easy axis of the specimen can guide the spin-flop transition in $Co_4Ta_2O_9$, whereas magnetic field deviating from the easy axis would steady the canted spin states. To estimate the critical magnetic field ($H_c$) for getting spin flop transition, we measured field dependent magnetization curve at 2 K for $Co_4Ta_2O_9$ (Fig. 3(b)). No sign of saturation magnetization (spin-flip phase) was found for magnetic fields up to 3 T. Moreover, a clear deviation in M is observed at $H_c$= 0.9 T that is clearly identified in dM/dH plot as seen in inset of Fig. 3 (b). This observation of increase in magnetization due to spin flop transition that occurs at such low fields implies that the magnetic lattice is extremely soft.[21] The important question therefore is what role this spin flop transition plays in achieving ME effects reported in $Co_4Ta_2O_9$.

To correlate the influence of underlying magnetic structure with electric ordering, next we discuss the temperature dependence of dielectric constant (ε-T) when the magnetic field is applied (Fig. 4 (a)). The room temperature electrical resistivity for $Co_4Ta_2O_9$ is estimated to be 1.57 x $10^7$ Ω-m, that implies that the specimen does not suffer from leakage problem. No deviation in dielectric constant across the Neel temperature is seen when the sample is cooled in zero external field. As the magnetic field is increased, a pronounced peak appears at the magnetic transition temperature ($T_N$=20 K) and the dielectric constant shows anomalous behaviour. From fig. 4 (a), we observe that the height of the peak increases with the magnetic field increment. Such type of magnetic field induced dielectric anomaly around $T_N$ has been observed in several other linear ME systems that is assigned to spin fluctuations.[5,11,12,14,22] Fig. 4(b) shows the magnetic field dependence of magnetocapacitance [MC(%) =[{ε(H) -ε(0)}/ ε (0) x 100] at different temperatures 15, 20 and 25 K, where ε (H) and ε (0) refer to dielectric constants in magnetic field (H) and zero fields respectively. Two important implications are observed at T = 20 K; (i) The magnitude of magnetocapacitance ($\Delta\varepsilon$ (6T)/ ε (0)) reaches 0.2% (at 10 kHz) and (ii) substantial hysteresis is observed. We note that such hysteretic behavior is not seen at 15 and 25 K.

Towards a qualitative understanding of our magneto-capacitance data in the presence of spin flop state, in the following we analyse our results in the framework of Ginzburg-Landau theory appropriate for a multiferroic system:

$$\phi = \phi_0 + \alpha P^2 + \beta/2\, P^4 - PE + \alpha' M^2 + \beta'/2 M^4 - MH + \gamma P^2 M^2 \ldots\ldots (1)$$

Where $\phi$, P and M are the thermodynamic potential, polarization and magnetization, respectively. $\phi_0$ is a reference potential and α, α′, β, β′ and γ are related coupling constants, respectively and the term $\gamma P^2 M^2$ shows the exchange magnetoelectric interaction term.[23] From, this general formalism, it can be shown that change in dielectric constant $\Delta\varepsilon \sim \gamma M^2$ where $\Delta\varepsilon$ is the change in dielectric constant which is proportional to square of magnetization. A linear relationship between $\Delta\varepsilon$ and $M^2$ gives the value of magnetoelectric coupling coefficient γ. Moreover, the sign of $\Delta\varepsilon$ depends upon the sign of coupling constant γ, that can be either positive or negative. To check the applicability of this theory, in inset of Fig. 4(b), magnetocapacitance is replotted as the fractional change of dielectric constant (magnetocapacitance [$\Delta\varepsilon/\varepsilon(0)$] vs $M^2$. It is evident that the magnetocapacitance vs. $M^2$ data at 20 K falls onto a single line. So the equation $\Delta\varepsilon \sim \gamma M^2$ broadly applies to $Co_4Ta_2O_9$. The M(H) values are obtained from the field dependence of magnetization (M-H) at fixed temperature (Fig 4(c)). The calculated value of γ from the inset of Fig. 4 (b) for $Co_4Ta_2O_9$ is found to be γ = 1.4 x$10^{-3}$ (emu/g)$^{-2}$ that confirms existence of large magnetoelectric coupling in $Co_4Ta_2O_9$.[24]

In Fig. 5 (a) we plot the temperature dependence of pyroelectric current for $Co_4Ta_2O_9$ when the compound was poled at different magnetic fields. The poling electric field was +200V/mm and it is evident that in the absence of magnetic field, no pyroelectric current is observed. On applying magnetic field, a pyroelectric current arises in the temperature interval below the AFM phase transition. The magnitude of current peak increases with increase in magnetic field. The temperature dependence of electric polarization (P-T), which is calculated by integrating the pyroelectric current with respect to time is shown in Fig. 5 (b). A maximum electric polarization of 52μC/m$^2$ is obtained at temperature 5 K and with magnetic field poling set to 6T. The inset of Fig. 5(b) shows the magnetic field dependence of polarization at 5, 10 and 15 K, respectively. The polarization increases with increasing magnetic field linearly upto 6 T, which again confirms the strong linear ME coupling.[10-12, 22] Fig. 5 (c) shows the change in sign of polarization at 3T magnetic field as a function of temperature with positive and negative poling electric fields (±200V/mm). A symmetric temperature dependence of polarization curve is observed, indicating that the electric polarization can be reversed by applying negative field. Generally, in case of linear ME materials, domain effects can effectively determine for the coupling nature.[25,26] Brown et al.[25] have analyzed such mechanism by measuring the relative populations of randomly oriented magnetic domains in $Cr_2O_3$. This process can be also useful to understand the origin of ME effect in $Co_4Ta_2O_9$ since the magnetic symmetry and crystal structure of this compound allows existence of the cross-coupling properties. In the absence of magnetic field, no ME signal is observed, which can be attributed to cancellation of ME coupling of the two different antiferromagnetic domains. When it undergoes Magnetic cooling, growth of one kind of the antiferromagnetic domains is preferred, which breaks the equilibrium between the two different types of domains resulting in net ME effect. In Fig. 6, we plot the pyroelectric current data with temperature oscillation in the antiferromagnetic state when the sample was cooled without magnetic field. We observe that the direction of pyroelectric current does not show any change during cooling and heating cycles. Clearly, when the temperature is varied from 18 K to 20 K in a periodic manner, there is no change in pyroelectric current. From Fig. 6, we also observe that the pyroelectric current does not show any change over time, in zero external magnetic field, which confirms the absence of genuine ferroelectricity in $Co_4Ta_2O_9$. This is the evidence for absence of spontaneous polarization state below the magnetic transition temperature.

## IV. Conclusions

In summary, our results on polycrystalline $Co_4Ta_2O_9$ (with space group $P\bar{3}c1$) reveal antiferromagnetic ordering at $T_N$=20 K and establish strong magneto-capacitance effect across the Neel temperature. Below $T_N$, temperature dependent magnetization curve exhibit spin-flop transition at 0.9 T that is linked to large magnetocapacitance of 0.2% (at 6 Tesla and 10 kHz). In accordance with Ginzburg-Landau formula for ferro-electromagnets, the change in dielectric constant is found to be proportional to square of magnetization and the magnetoelectric coupling parameter is estimated to be $1.4\times10^{-3}$ $(emu/g)^{-2}$. Linear ME effect and polarization reversal are also confirmed from pyroelectric current data. Temperature oscillation measurements indicate that the ground state of $Co_4Ta_2O_9$ is non-ferroelectric.

**Acknowledgements**

S. Chaudhary acknowledges UGC, New Delhi-India for Fellowship. P. Srivastava acknowledges UGC, New Delhi-India for Dr. D. S. Kothari Post Doctoral Fellowship. We thank AIRF- JNU for SEM and magnetization measurements. SP thanks funding support of DST- SERB, DST-FIST and DST-PURSE programs of Government of India.

**Figure Captions:**

**Table 1.** Atomic coordinates derived from crystal structure analysis of $Co_4Ta_2O_9$.

**Fig.1** Room temperature powder XRD pattern of polycrystalline $Co_4Ta_2O_9$ sample with Reitveld refinement. The observed data (red), calculated line (black) and difference (blue) between the two are shown along with the Bragg positions (green). **(b)** Schematic crystal structure of $Co_4Ta_2O_9$.

**Fig. 2 (a)** SEM image; **(b)** EDX pattern of $Co_4Ta_2O_9$.

**Fig. 3(a)** The temperature dependence of magnetization is plotted at 0.01, 0.1, 1, 1.5, 2 T, respectively; inset shows temperature dependence of dM/dT; **(b)** The magnetization (M) versus magnetic field (H) at 2 K for $Co_4Ta_2O_9$; The inset shows the dM/dH as a function of magnetic field.

**Fig.4 (a)** Dielectric constant is plotted as a function of temperature under various magnetic fields; The data were measured at 10 KHz. **(b)** Magnetocapacitance as a function of magnetic field at various temperatures (15, 20 and 25 K); the inset shows a linear relationship between the measured magnetocapacitance plotted against square of magnetization ($M^2$) at 20 K for $Co_4Ta_2O_9$; **(c)** Field dependence of magnetization measured at 15 K, 20 K and 25 K.

**Fig.5 (a)** Pyroelectric current as a function of temperature under various magnetic fields; **(b)** Calculated electric polarization as a function of temperature at various magnetic fields; The inset shows the polarization versus magnetic field at 5, 10 and 15 K, respectively. **(c)** Electric polarization is plotted as a function of temperature at various poling field. The direction of polarization reversal is also confirmed for negative poling field.

**Fig. 6.** Pyroelectric current as a function of time (t) with thermal cycle in the antiferromagnetic state.

**Table 1**

| Atom | X | Y | Z |
|---|---|---|---|
| Ta1 | 0.00000 | 0.00000 | 0.14284 |
| Co1 | 0.33330 | 0.66660 | 0.01900 |
| Co2 | 0.33330 | 0.66660 | 0.30340 |
| O1 | 0.30600 | 0.30800 | 0.08600 |
| O2 | 0.27600 | 0.00000 | 0.08300 |

**Table 1**

**Fig. 1**

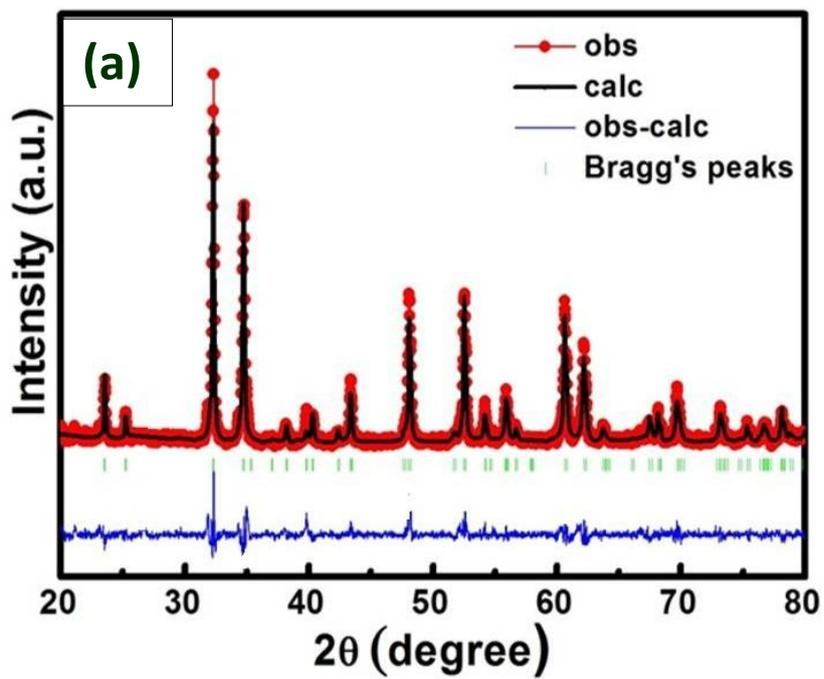

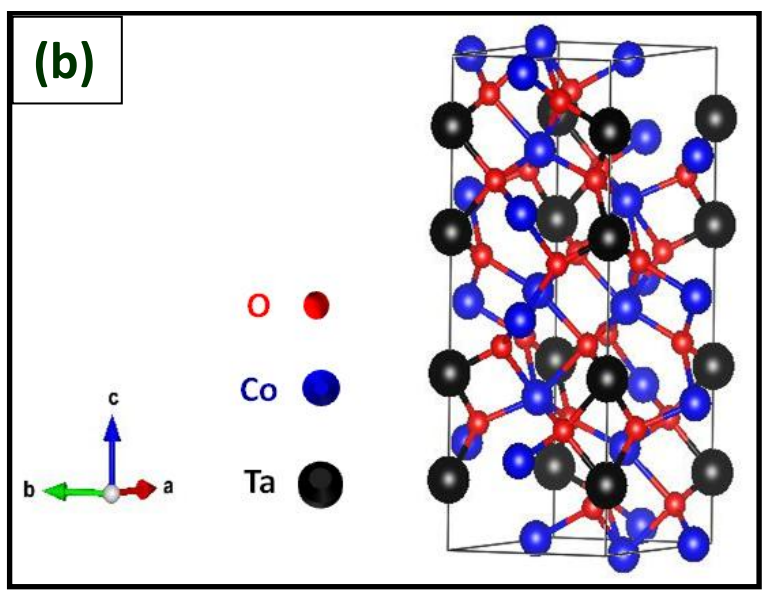

**Fig. 2**

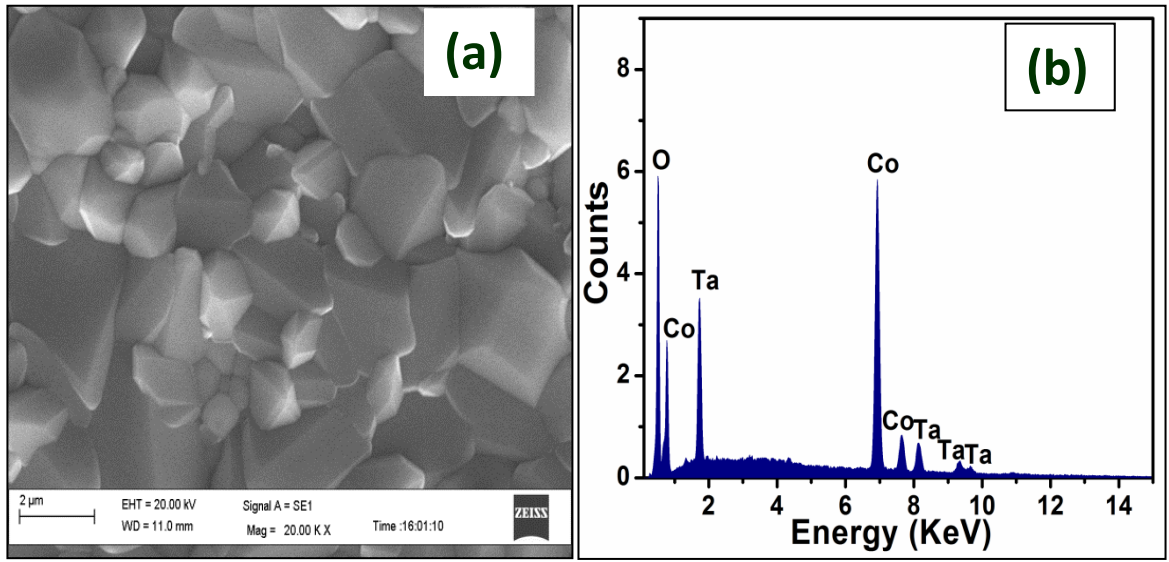

**Fig. 3**

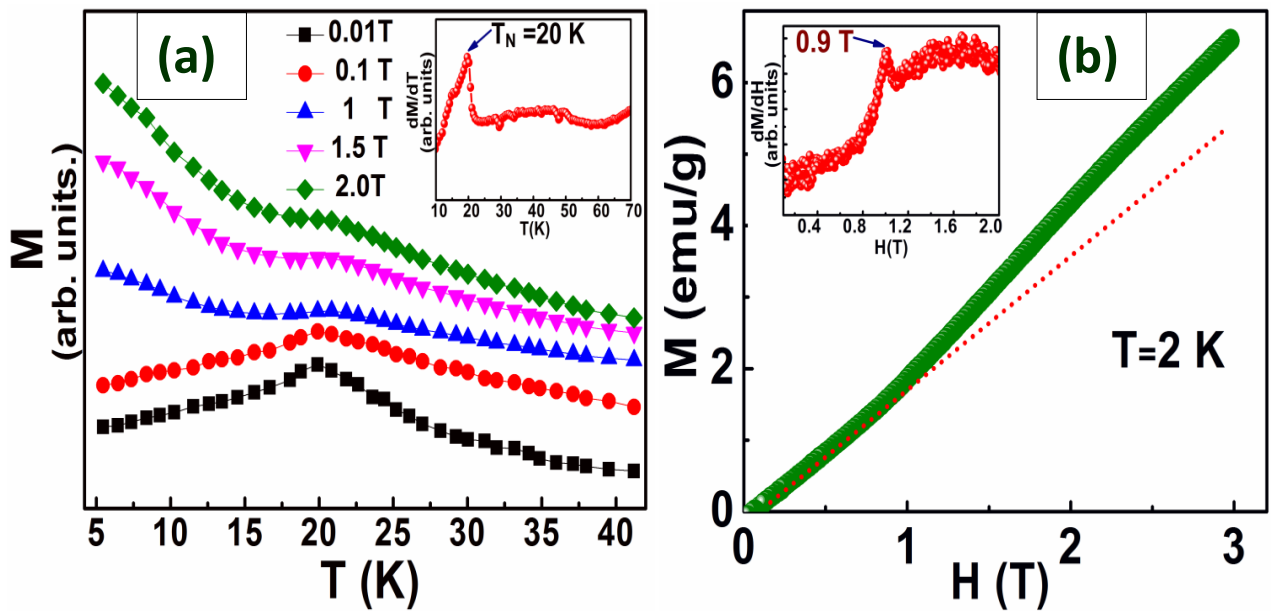

**Fig. 4**

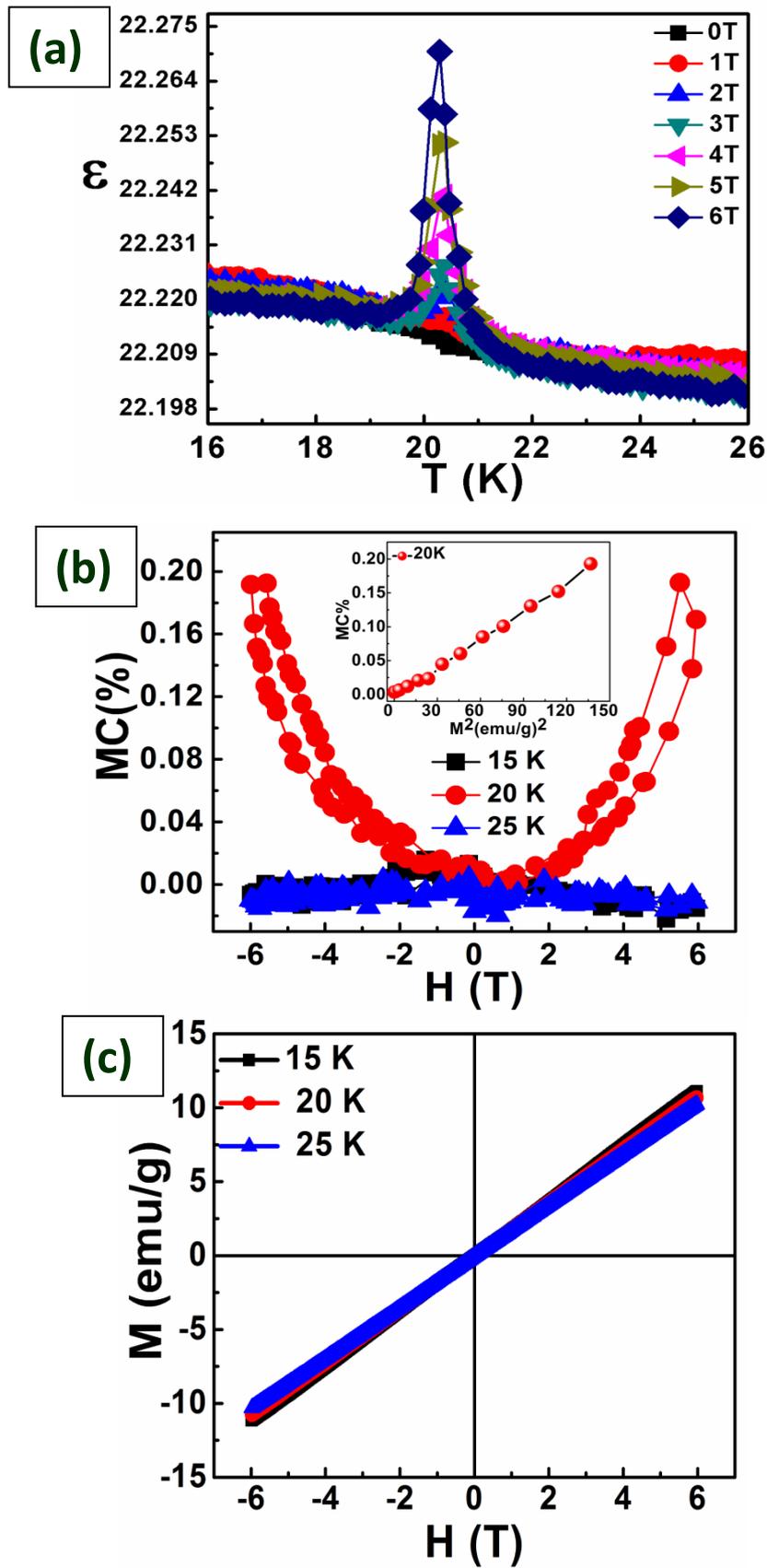

**Fig. 5**

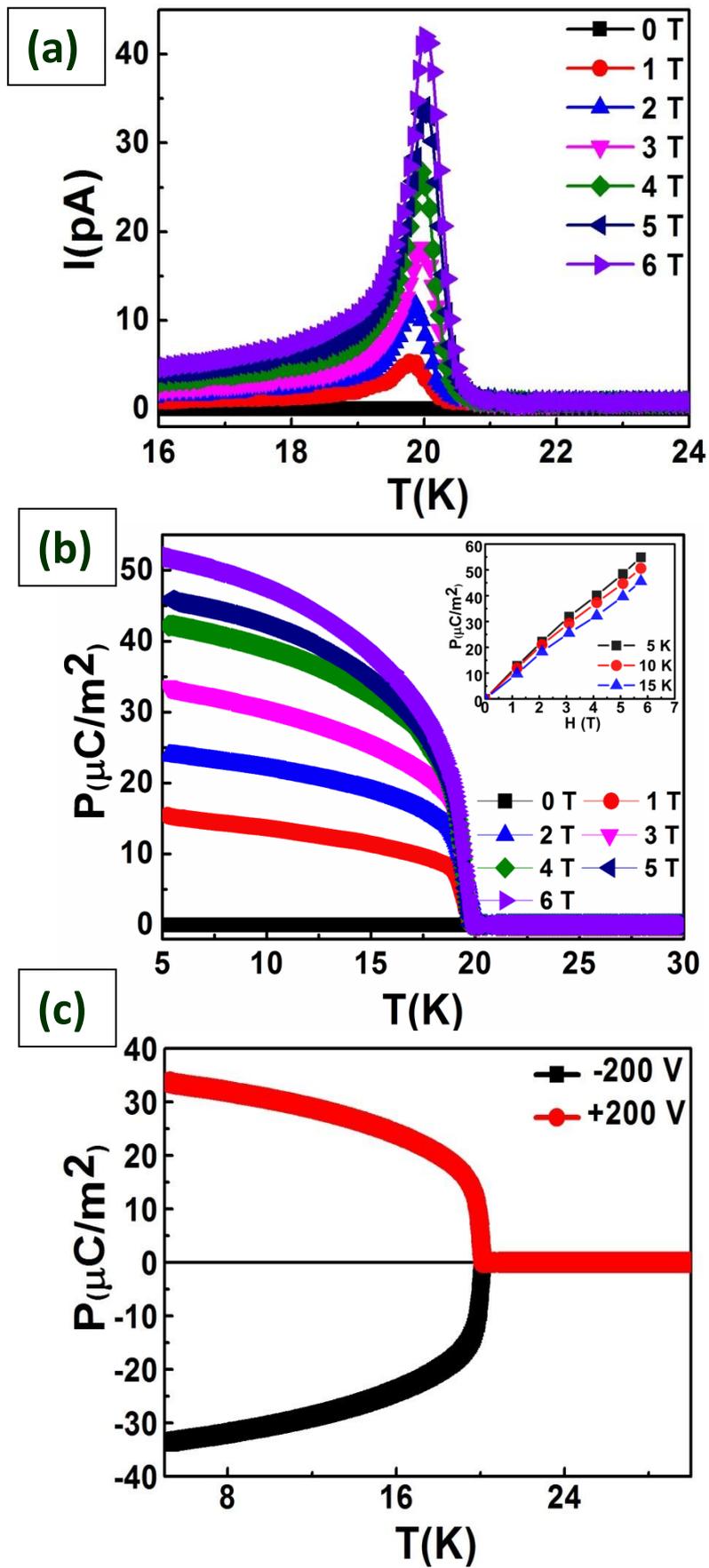

**Fig. 6**

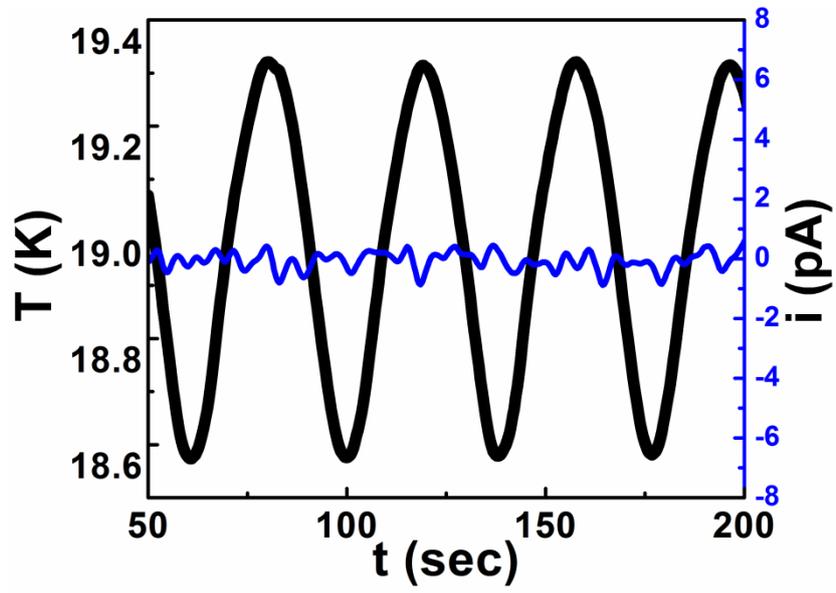